\documentclass[twocolumn,preprintnumbers,amsmath,amssymb,superscriptaddress, altaffilletter, subeqn,prl]{revtex4-1}

\usepackage{graphicx}
\usepackage{times}
\usepackage{soul}
\usepackage{color} 
\usepackage{mathtools}
\usepackage{siunitx}
\usepackage{comment}
\usepackage{mathrsfs}

\def\bra#1{\mathinner{\langle{#1}|}}
\def\ket#1{\mathinner{|{#1}\rangle}}

\newlength{\singlecolumn}
\renewcommand{\vec}[1]{\mathbf{#1}}

\makeatletter
\renewcommand*{\@fnsymbol}[1]{\ensuremath{\ifcase#1\or \dagger\or *\or  \ddagger\or
\mathsection\or \mathparagraph\or \|\or **\or \dagger\dagger
\or \ddagger\ddagger \else\@ctrerr\fi}}
\makeatother

\begin{document}

\title{Signatures of Dimensionality and Symmetry in Exciton Bandstructure: \\Consequences for Time-Evolution}

\author{Diana Y. Qiu*}
\email{diana.qiu@yale.edu}
\affiliation{Department of Mechanical Engineering and Materials Science, Yale University New Haven, CT 06516, USA}
\author{Galit Cohen}
\affiliation{Department of Materials and Interfaces, Weizmann Institute of Science, Rehovot 7610001, Israel}
\author{Dana Novichkova}
\affiliation{Department of Materials and Interfaces, Weizmann Institute of Science, Rehovot 7610001, Israel}
\author{Sivan Refaely-Abramson*}
\email{sivan.refaely-abramson@weizmann.ac.il}
\affiliation{Department of Materials and Interfaces, Weizmann Institute of Science, Rehovot 7610001, Israel}

\begin{abstract}
   Exciton dynamics, lifetimes and scattering are directly related to the exciton dispersion, or bandstructure. While electron and phonon bandstructures are well understood and can be easily calculated from first principles, the exciton bandstructure is commonly conflated with the underlying electronic bandstructure, where the exciton dispersion is assumed to follow the same dispersion as the electron and hole bands from which it is composed (i.e., the effective mass model). Here, we present a general theory of exciton bandstructure within both \emph{ab initio} and model Hamiltonian approaches. We show that contrary to common assumption, the exciton bandstructure contains non-analytical discontinuities---a feature which is impossible to obtain from the electronic bandstructure alone. These discontinuities are purely quantum phenomena, arising from the exchange scattering of electron-hole pairs. We show that the degree of these discontinuities depends on materials' symmetry and dimensionality, with jump discontinuities occurring in 3D and different orders of removable discontinuities in 2D and 1D.
   We connect these unexpected features to the early stages of exciton dynamics, which shows remarkable correspondence with recent experimental observations and suggests that the measured diffusion patterns are influenced by the underlying exciton bandstructure. 
\end{abstract}

\maketitle

In low-dimensional and nanostructured materials, the optical response is dominated by correlated electron-hole pairs---or excitons---bound together by the Coulomb interaction. 
By now, it is well-established that these large excitonic effects are a combined consequence of quantum confinement and reduced screening in low dimensions~\cite{spataru2004a, deslippe2009, Qiu2013, Qiu2016, Chernikov2014, Wang2018}. However, many challenges remain in understanding the time evolution of these excitons, especially when it comes to correlating complex experimental signatures with underlying physical phenomena through the use of quantitatively predictive theories.

Recent advances in temporally and spatially-resolved microscopies, such as transient absorption microscopy and time-resolved photoluminesence, allow direct observation of exciton relaxation and diffusion processes~\cite{Ginsberg2020, Zhu2019, amori2018}. For example, exciton diffusion in mono- and few-layer transition metal dichalcogenides (TMDs) reveal a ring-like diffusion pattern~\cite{Kulig2018}, while acene molecular crystals exhibit an asymmetric exciton diffusion with signatures of interconversion between exciton singlet and triplet states~\cite{Wan2015, schnedermann2019}. These diffusion processes reveal a wealth of competing scattering mechanisms and decay pathways, involving exciton-phonon interactions~\cite{kato2016, Ginsberg2020, glazov2019}, exciton-exciton annihilation~\cite{Kumar2014, Mouri2014, Sun2014}, occupation of dark states~\cite{ Yuan2017,  Wan2015, Yong2017}, electron-hole plasma~\cite{zipfel2020, sternbach2020}, and environmental disorder~\cite{penwell2017, Delor2020, zipfel2020}.

From a theoretical point of view, rate-equation approaches are typically used to model exciton diffusion, but these demand extensive parametrization~\cite{Wan2015, perea2019}, which hinders a predictive  understanding of exciton structure-property relations. For near-equilibrium excitons in the low-field limit, \textit{ab initio} Green's function-based many-body perturbation theory, within the GW plus Bethe Salpeter equation (GW-BSE) approach \cite{Hedin1965, Hedin1969, Hybertsen1985, Hybertsen1986, Albrecht1998, Shirley1998, Rohlfing1998, Rohlfing2000, Deslippe2012}, has been highly successful in predicting optical spectra across a wide variety of materials of different dimensionalities~\cite{louie2005, xie2019}. GW-BSE methods can also give insight into exciton dynamics, including radiative recombination processes~\cite{Spataru2005, Palummo2015, gao2017, Chen2018, chen2019}, multiexciton generation~\cite{deilmann2016, Refaely-Abramson2017, arora2019}, and exciton-phonon interactions~\cite{Antonius2017,Chen2020, Alvertis2020}. 

Traditional implementations of the GW-BSE method focus on zero-momentum excitons, but exciton dynamics require knowledge of the exciton bandstructure to accurately describe the phase space of momentum-conserving scattering processes. Until recently, the vast majority of theoretical approaches assumed that the exciton dispersion is either non-dispersive~\cite{Frenkel1931} or free-particle like---i.e. parabolic with an effective mass equivalent to the sum of the quasielectron and quasihole masses~\cite{Wannier1937, haug2009}. However, recent \textit{ab initio} GW-BSE calculations of exciton bandstructure reveal that v-shaped, nonanalytic dispersion can arise in 2D materials~\cite{Qiu2015, Cudazzo2016} and bands of both positive and negative mass appear in the bandstructure of acene molecular crystals~\cite{Cudazzo2015, Refaely-Abramson2017}. Nonetheless, a general comprehensive understanding of how crystal symmetry and dimensionality affects exciton bandstructure and how signatures of that bandstructure manifest in exciton dynamics is still needed.

In this work, we explore signatures of symmetry and dimensional confinement in exciton bandstructures and their manifestation thereof in the exciton time-evolution within the coherent, quasi-ballistic regime immediately following excitation.  We use the GW-BSE method, modified for finite momentum, to compute exciton bandstructures of four prototypical systems of different dimensionality and symmetry.
Remarkably, we find that contrary to common assumption, the exciton bandstructure contains non-analytical discontinuities at $\Gamma$ in materials of all dimensions---a feature that is impossible to obtain from the electronic bandstructure alone. These discontinuities are a purely quantum phenomenon, arising from the exchange scattering of electron-hole pairs, and the degree of these discontinuities depends on the materials' symmetry and dimensionality, with jump discontinuities occurring in 3D and different orders of removable discontinuities in 2D and 1D.
Finally, we show that discontinuities in the exciton bandstructure can manifest in unexpected nodal structure in the time-evolution of an initially Gaussian exciton wavepacket, resulting in intriguing qualitative similarities to recent experiments in the diffusive regime, which suggest that exciton bandstructure may play a previously overlooked role in establishing the initial conditions of diffusion.

\begin{figure*}
\begin{center}
\includegraphics[width=\textwidth]{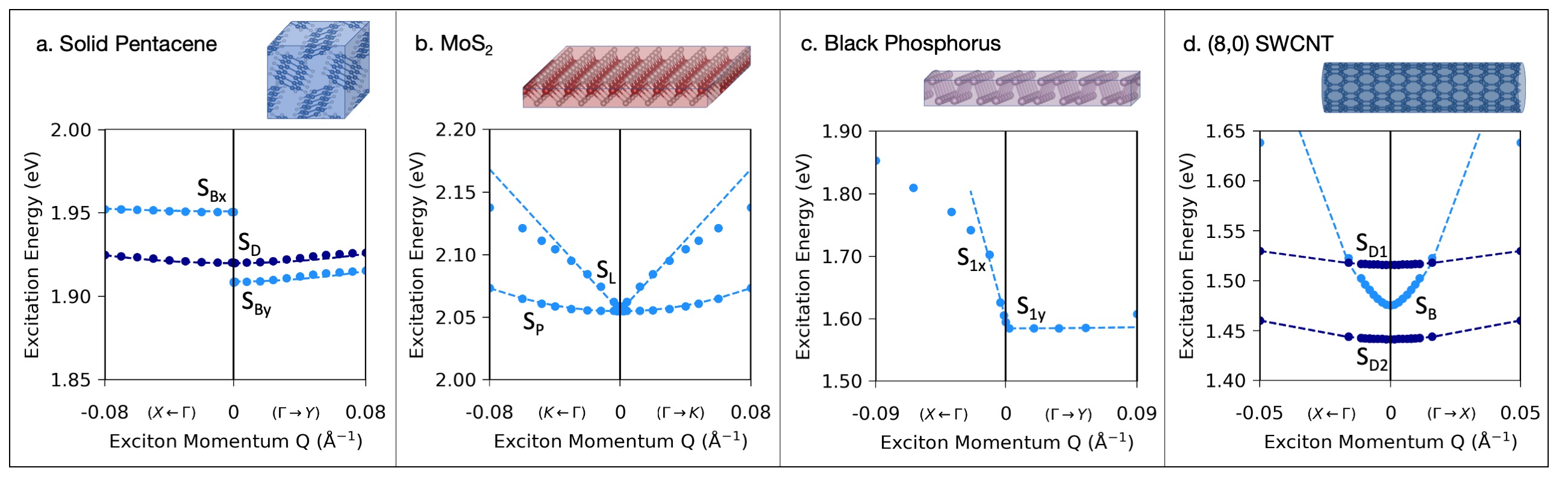}
\caption{Exciton bandstructures for spin S=0 excitons in different materials. Dots are GW-BSE results. Dashed lines show the fit to a model Hamiltonian. (a) Solid pentacene, showing two nearly-degenerate singlet exciton states that are dipole-allowed (S$_B$, light blue) and dipole-forbidden (S$_D$, dark blue)  at $\textbf{Q}=0$;  (b) monolayer MoS$_2$, showing two exciton bands degenerate at $\textbf{Q}=0$, with linear (S$_L$) and parabolic (S$_P$) dispersion;  (c) monolayer black phosphorus, with linear dichroism manifesting in a dipole-allowed (bright) exciton at $\textbf{Q}=0$ (S$_B$) with linear dispersion along the $\Gamma$ to $X$ direction; and (d) (8,0) single-walled carbon nanotube, showing a non-parabolic low-lying exciton that is dipole-allowed at $\textbf{Q}=0$ (S$_{B}$, light blue) and two nearby dark parabolic exciton bands (S$_{D1}$ and S$_{D2}$, dark blue).}
\label{fig:Fig1}
\end{center}
\end{figure*}

We begin by calculating the exciton bandstructure for four exemplary materials of different dimensionality and current interest: 1) the pentacene molecular crystal, a three-dimensional (3D) system with large excitonic effects~\cite{Sharifzadeh2012,Sharifzadeh2013}; 2)  monolayer MoS$_2$, a quasi-two-dimensional (quasi-2D) system with degenerate excitons in different momentum-space valleys~\cite{Qiu2013,Qiu2016}; 3) 2D black phosphorus, a quasi-2D system with strongly-bound excitons exhibiting linear dichroism~\cite{Wang2015,Qiu2017}; and 4) the $(8,0)$ single-walled carbon nanotube, a prototypical quasi-one-dimensional (quasi-1D) system known to host large excitonic effects~\cite{spataru2004a,spataru2004b,Spataru2005}.  Fig.~\ref{fig:Fig1} shows the calculated GW-BSE exciton bandstructure near $\textbf{Q}=0$ for these four materials (see SI for computational details). We see that in all cases "bright" (i.e., dipole allowed) exciton bands (light blue points in Fig.~\ref{fig:Fig1}) exhibit non-analytic behavior at the $\Gamma$ point in the exciton bandstructure, while "dark" exciton bands for which the transition from the ground state is dipole forbidden are parabolic (dark blue points in Fig.~\ref{fig:Fig1}). For simplicity, Fig.~\ref{fig:Fig1} only shows the spin S=0 (or spin-singlet) excitons. The spin S=1 (or spin-triplet) excitons have no exchange contribution and thus have a parabolic dispersion as we explain below.

To understand the source of the non-analytic dispersion, we derive a model Hamiltonian fit to our \textit{ab initio} results. We start with the expression for the \textit{ab initio} BSE. In our calculations, electron-hole interactions are built on top of the quasiparticle (QP) picture by solving the BSE in the electron-hole basis~\cite{Strinati1988,Rohlfing2000,Deslippe2012}: 
\begin{equation}
  \label{eq:bse}
\begin{split}
  \left(E_{c\textbf{k+\textbf{Q}}}-E_{v\textbf{k}}\right)A_{vc\textbf{k}\textbf{Q}}^{S}
  \\ +\sum_{v'c'\textbf{k}'}
   \bra{v\vec{k};c\textbf{k}+\textbf{Q}}&K^{eh}\ket{v'\textbf{k}';c'\textbf{k}'+
   \textbf{Q}}A_{v'c'\textbf{k}'\textbf{Q}}^{S}
  \\ =\Omega^{S}_\textbf{Q} A_{vc\textbf{k}\textbf{Q}}^{S}.
  \end{split}
\end{equation}
Here, the index $(v\textbf{k};c\textbf{k}+\textbf{Q})$ indicates a hole state $|v\textbf{k}\rangle$ and an electron state $|c\textbf{k}+\textbf{Q}\rangle$, where $\textbf{k}$ is the crystal momentum and $\textbf{Q}$  is the exciton center-of-mass momentum; $E_{c\vec{k+\vec{Q}}}$ and $E_{v\vec{k}}$ are the QP energies calculated within the GW approximation;  $S$ indexes the exciton state at momentum $\textbf{Q}$; $A_{v\vec{k};c\vec{k}+\vec{Q}}^{S}$ is the amplitude of the free electron-hole pair; 
$\Omega^S_\textbf{Q}$ is the exciton excitation energy;  and $K^{eh}$ is the electron-hole interaction kernel. To obtain the exciton bandstructure (or dispersion), we solve the BSE at different exciton momenta $\textbf{Q}$ following the methodology developed in Ref.~\cite{Gatti2013} and ~\cite{Qiu2015}.

The first term in Eq.~\ref{eq:bse}, $ (E_{c\vec{k+\vec{Q}}}-E_{v\vec{k}})$, is the electron-hole transition energy, analogous to the kinetic energy. When the electron-hole interaction is small, this term dominates, and as long as the electron and hole each arise from a single parabolic band, the exciton dispersion is parabolic with the mass of the exciton equal to the sum of the electron and hole band masses. This is the limit where the commonly used effective mass approximation holds. The interaction kernel in Eq.~\ref{eq:bse} is $K^{eh}=K^d+K^x$ (or $K^{eh}=K^d+2K^x$, when the spin-orbit interaction is neglected). It consists of a direct term ($K^d$), where the attractive interaction between the electron and the hole is mediated by the screened Coulomb interaction,
\begin{equation}
\begin{split}
\label{eq:Kd}
   & \bra{v\textbf{k};c\textbf{k}+\textbf{Q}}K^{d}\ket{v'\textbf{k}';c'\textbf{k}'+\textbf{Q}} \\
    &= -\sum_{\textbf{GG}'}M^*_{cc'}(\textbf{k}+\textbf{Q},\textbf{q},\textbf{G})W_{\textbf{GG}'}(\textbf{q})M_{vv'}(\textbf{k},\textbf{q},\textbf{G'}),
    \end{split}
\end{equation}
and an exchange term ($K^x$), where the exchange scattering of an electron-hole pair gives rise to a repulsive term mediated by the bare Coulomb interaction,
\begin{equation}
\begin{split}
\label{eq:Kx}
& \bra{v\textbf{k};c\textbf{k}+\textbf{Q}}K^{x}\ket{v'\textbf{k}';c'\textbf{k}'+\textbf{Q}} \\
& = \sum_{\textbf{G}}M^*_{cv}(\textbf{k},\textbf{Q},\textbf{G})\mathrm{v}(\textbf{Q}+\textbf{G})M_{c'v'}(\textbf{k}',\textbf{Q},\textbf{G}).
\end{split}
\end{equation}
Here, $W$ and $\mathrm{v}$ are the screened and the bare Coulomb interactions, respectively; $\textbf{G}$ are the reciprocal lattice vectors; $\textbf{q}=\textbf{k}-\textbf{k}'$; and $M$ are the plane-wave matrix elements such that $M_{nn'}(\textbf{k},\textbf{q},\textbf{G})=\langle n\textbf{k}|e^{i(\textbf{q}+\textbf{G})\cdot\textbf{r}}|n'\textbf{k}'\rangle$ \cite{Rohlfing1998, Deslippe2012}. It is important to note that in typical calculations of optical spectra within the BSE, the long-range (i.e. $\textbf{G}=0$) term in Eq.~\ref{eq:Kx} is neglected~\cite{Onida2002,Gatti2013,Deslippe2012} in order to avoid the non-analytic behavior of $\mathrm{v}$. This is equivalent to solving for only transverse excitons at exactly $\textbf{Q}=0$~\cite{Denisov1973} but leads to the introduction of spurious results at finite momentum (see SI). Thus, we solve the BSE including both long and short-range exchange, and the resulting exciton bandstructure contains \textit{both} longitudinal and transverse solutions.

We are interested in the exciton dispersion near the band edge at $\textbf{Q}=0$. In the limit of small $\textbf{Q}$ and $\textbf{G}=0$ (the long wavelength limit), the direct term goes as $K^d (\textbf{Q})\propto-Q^2$, where the $Q^2$  dependence comes from the limiting behavior of the $M$ matrix elements and the negative sign comes from the attractive Coulomb potential and results in an enhancement of the exciton effective mass~\cite{Qiu2015}. The exchange term goes as 
\begin{equation}
\label{eq:Kxlimit}
K^x(\textbf{Q})\propto (\textbf{Q}\cdot\textbf{p})^2\mathrm{v}(\textbf{Q}),
\end{equation}
where $\textbf{p}=\langle 0|\textbf{r}|S(\textbf{Q})\rangle$ is the dipole matrix element of the exciton state $|S(\textbf{Q})\rangle$ (see SI). The Coulomb interaction in reciprocal space approaches different limits depending on the dimensionality (see SI),  
leading to different dimension-dependent small $\textbf{Q}$ (or long wavelength) limits of the exchange:
\begin{equation}
    K^x(\textbf{Q}\rightarrow 0)\propto
    \begin{matrix}
    \cos^2(\theta_\textbf{Q})\times\mathrm{constant} & & \mathrm{in \, 3D\quad} \\
    |Q|\cos^2(\theta_\textbf{Q}) & & \mathrm{in \, 2D \quad} \\
    -2Q^2\cos^2(\theta_\textbf{Q})(\gamma_E + \ln||) & & \mathrm{in \, 1D\quad} \\
    \end{matrix},
    \label{eq:Kx-dim}
\end{equation}
where $\gamma_E=0.577$ is the Euler constant~\cite{mihaila2011}, and $\theta_\textbf{Q}$ is the angle between the exciton momentum, $\textbf{Q}$, and the matrix element, $\textbf{p}$. Thus, the behavior of the exchange term can vary dramatically with the dimensionality of the system under consideration and the anisotropy of the exciton's dipole matrix element.
The long wavelength component of the exchange term discussed above is only non-zero for excitons with a longitudinal component (i.e. $\textbf{Q}\not\perp \textbf{p}$)~\cite{Denisov1973}. The longitudinal excitons correspond to those measured by the energy-loss function~\cite{Gatti2013}, and may also be optically active in low-dimensional materials--when the polarization of the light has a component parallel to the momentum transfer in the crystal's periodic direction~\cite{Qiu2015}--and in bulk materials though interactions with the degenerate transverse polariton~\cite{Andreani1988}.

Following Eq.~\ref{eq:Kx-dim}, we can now derive a general form of the exciton dispersion in all dimensions (see SI for the full model Hamiltonian). In 3D, the exciton dispersion has the general form 
\begin{equation}
\label{eq:3D_dispersion}
    \Omega(\textbf{Q})=\Omega_0 + C\cos^2(\theta_\textbf{Q}) + \frac{\hbar^2}{2}(\frac{Q_x^2}{M^*_x}+\frac{Q_y^2}{M^*_y}+\frac{Q_z^2}{M^*_z}),
\end{equation}
where $\Omega_0$ is the excitation energy of the dipole forbidden exciton; $\Omega_0+C$ is the excitation energy of the dipole-allowed exciton; $\theta_\textbf{Q}$ is the angle of $\textbf{Q}$ with respect to $\textbf{p}$; and $M^*$ is the exciton's effective mass along each of the three Cartesian directions $x, y, z$ (see SI). 

With this understanding, we can interpret the dispersion of the pentacene crystal (Fig.~\ref{fig:Fig1}a).
The exciton bandstructure shows two low-lying  optically bright ($S_B$) and dark ($S_D$) spin-singlet states~\cite{Refaely-Abramson2017, Cudazzo2015}, with significant mass enhancement compared to the underlying quasiparticle band masses (see SI). Additionally, the $S_B$ band exhibits a discontinuous jump at $\textbf{Q}=0$, which arises due to the asymmetry of the crystal. The absorption of light is only dipole allowed for polarizations along the $\textbf{a}$-axis, namely $\textbf{p}$ of the bright exciton is oriented along the $(\Gamma\rightarrow X)$ direction. Thus, the longitudinal exciton, for which $\theta_\textbf{Q}=0$, has momenta along $(\Gamma\rightarrow X)$, while the transverse exciton, for which $\theta_\textbf{Q}=90^{\circ}$ has momenta along $(\Gamma\rightarrow Y)$. The energy of the longitudinal branch is increased by a constant, $C$ in Eq.~\ref{eq:3D_dispersion}, due to the long-range limit of the exchange $K^x$, resulting in the discontinuity at $\textbf{Q}=0$.
This discontinuity has also been observed in previous \textit{ab initio} calculations of exciton dispersion in molecular crystals, and was associated to Davydov splitting~\cite{Cudazzo2015}. We note that this splitting, when considering the long wavelength limit, is a general phenomenon for optically bright excitons in 3D crystals exhibiting anisotropic light absorption and can be understood purely from the selection rules of the Bloch states, independent of simplifying assumptions about the Wannier or Frenkel nature of the excitons.
On the other hand, the dark singlet state $S_D$ exhibits no such discontinuity because it is composed of dipole-forbidden electron-hole transitions, and consequently, the long-range exchange term is vanishingly small.

Following Eq.~\ref{eq:Kx-dim}, in 2D, the exciton dispersion of a non-degenerate exciton has the general form  
\begin{equation}
\label{eq:2d_dispersion}
    \Omega(\textbf{Q})=\Omega_0 + A|Q|\cos^2(\theta_\textbf{Q})+ \frac{\hbar^2Q_x^2}{2M_x^*} + \frac{\hbar^2Q_y^2}{2M_y^*} ,
\end{equation}
where the $|Q|$ term comes from the momentum-dependent behavior of the exchange interaction in 2D and can give rise to a linear dispersion that is non-analytic at $\textbf{Q}=0$ (see SI). This general form is reflected in the calculated exciton dispersion of monolayer black phosphorus (Fig.~\ref{fig:Fig1}c). In black phosphorus, for the lowest energy singlet exciton, the dispersion is linear along the $\Gamma$ to $X$ direction, while the dispersion is parabolic along the $\Gamma$ to $Y$ direction. This is a consequence of the crystalline asymmetry: optical absorption is only allowed for polarizations along the armchair (or $\Gamma$ to $X$) direction, and the dipole matrix element of the bright exciton is likewise oriented along $\Gamma$ to $X$. Thus, the long-range exchange interaction, which gives rise to the $|Q|$ dispersion, is zero for excitons with center-of-mass momentum, $\textbf{Q}$, along the zigzag (or $\Gamma$ to $Y$) direction perpendicular to the exciton dipole matrix element but goes as $|{Q}|$ for excitons with center-of-mass momentum, $\textbf{Q}$ parallel to the exciton dipole matrix element ($\Gamma$ to $X$). 

In monolayer MoS$_2$ (Fig.~\ref{fig:Fig1}b), the picture changes slightly due to the two-fold degeneracy of the lowest energy exciton at $\textbf{Q}=0$. As previously reported in Ref.~\cite{Qiu2015}, the interplay of intervalley and intravalley exchange in MoS$_2$ gives rise to two optically-bright low-energy exciton bands that are degenerate at $\textbf{Q}=0$. In the upper band  the intervalley and intravalley exchange add coherently, leading to a massless v-shaped dispersion, while in the lower band the intervalley and intravalley exchange cancel, leaving a parabolic dispersion whose effective mass is enhanced by the large exciton binding energy. In the small $\textbf{Q}$ limit, the dispersion of the linear band ($S_L$) to lowest order in $\textbf{Q}$ follows
\begin{equation}
\label{eq:mos2_dispersionL}
    \Omega(\textbf{Q})=\Omega_0 + 2A|Q|,
\end{equation}
where $A$ is a constant. The parabolic band ($S_P$) follows
\begin{equation}
\label{eq:mos2_dispersionP}
    \Omega(\textbf{Q})=\Omega_0 + \frac{\hbar^2Q^2}{2M^*}.
\end{equation}
 We thus see that in quasi-2D the long-range limit of the exchange in the electron-hole basis (Eq.~\ref{eq:Kx}) allows for the possibility of linear dispersion, while the dipole selection rules along different directions, along with the excitonic mixing of different valleys, determines how a given excitonic band will disperse independent of the dispersion of the underlying quasiparticle bands. Unlike in 3D, there is no splitting of the longitudinal and transverse modes in quasi-2D at $\textbf{Q}=0$, and the purely longitudinal mode can be optically active, since the polarization of the external field is not constrained to be parallel to the exciton dipole matrix element (see SI).

In 1D, following Eq.~\ref{eq:Kx-dim}, the exciton dispersion has the general form 
\begin{equation}
    \label{eq:swcnt_dispersion}
    \Omega(\textbf{Q})=\Omega_0 + B Q^2\ln{|Q|}
    + \frac{\hbar^2Q^2}{2M^*} ,
\end{equation}
where now the long-range exchange introduces a term that goes as $Q^2\ln{|Q|}$. This behavior is reflected in the dispersion of excitons in the (8,0) SWCNT ( Fig.~\ref{fig:Fig1}d). This quasi-1D system contains a large number of strongly-bound spin-singlet excitons ~\cite{spataru2004a, spataru2004b}, with several optically-dark states below and around the first bright peak at $\sim$1.5 eV~\cite{Spataru2005}. For simplicity, we focus on the exciton states close in energy to the lowest energy bright exciton. Here, the quasi-1D confinement leads to two types of very different low-lying singlet states. The dispersion of the optically bright state $S_{B}$ fits well to the expression in Eq.~\ref{eq:swcnt_dispersion}, with B$\sim$50. In contrast, the two nearby dark states, $S_{D_{1,2}}$, which have vanishingly small long-range exchange interactions, have a parabolic shape with relatively large effective masses, namely B=0 (see SI). In 1D, as in 2D, there is no splitting of the longitudinal and transverse modes. Since the polarization of the external field is not constrained to be parallel to the exciton dipole matrix element, the exchange contributes to both longitudinal and transverse modes, and the purely longitudinal mode can be optically active, as is the case in the $(8,0)$ SWCNT.

\begin{figure*}
\begin{center}
\includegraphics[width=\textwidth]{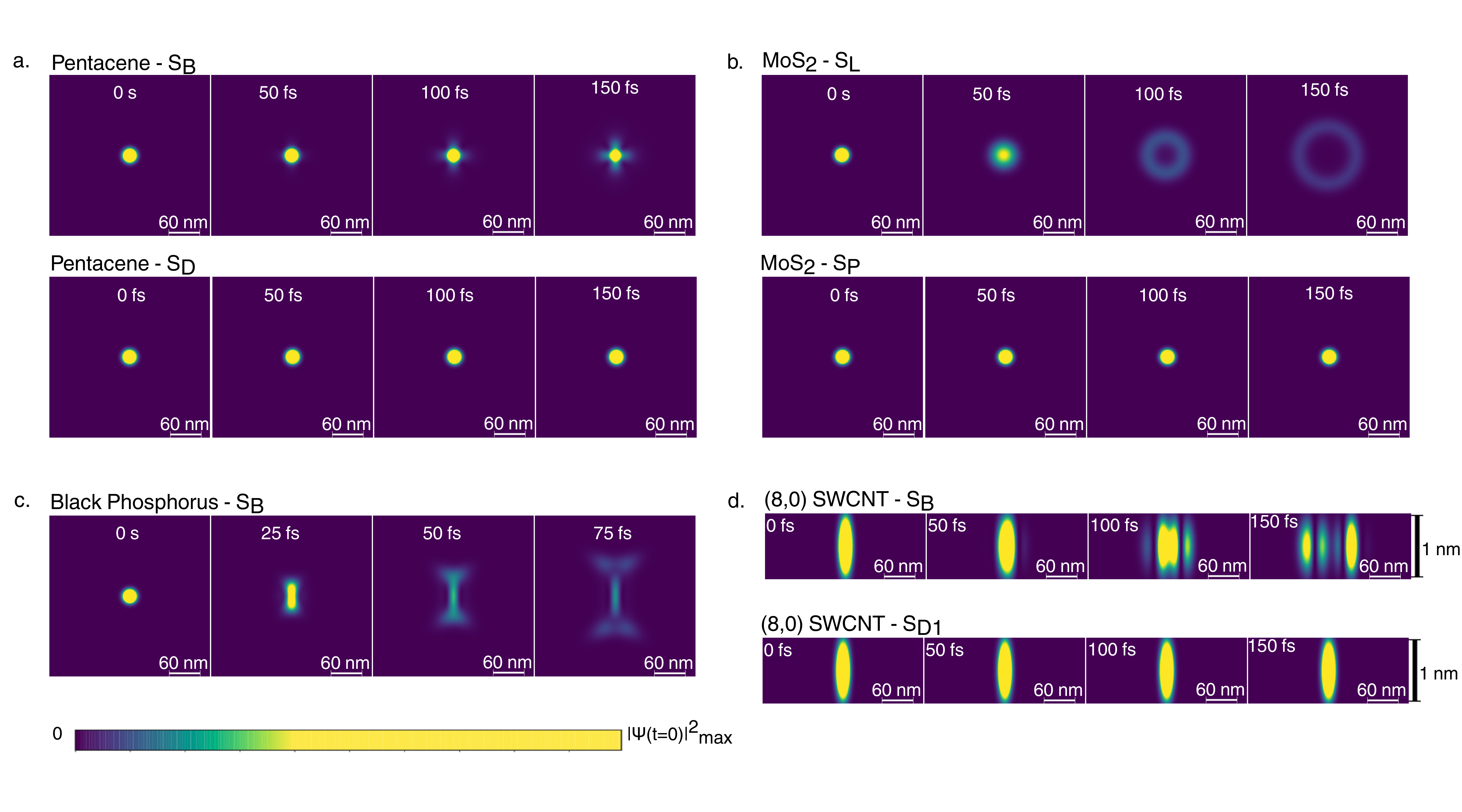}
\caption{Time evolution of the amplitude squared ($|\Psi(\mathbf{R},t)|^2$) of an initial Gaussian exciton wavepacket as a function of the bandstructure of a single exciton band, as labeled in Fig.~\ref{fig:Fig1}, for a) pentacene, b) monolayer MoS$_2$, c) monolayer black phosphorus, and d) the (8,0) SWCNT. The initial wavepacket at $t=0$ has the same spatial distribution in all cases, but the different band dispersion leads to distinct features in the wavepacket evolution with time. In all cases, the distribution is normalized with respect to the maximum value at time $t=0$.}.
\label{fig:Fig2}
\end{center}
\end{figure*}

Finally, we examine how signatures of symmetry and dimensionality in the exciton bandstructure manifest in the exciton dynamics. Here, we consider the semiclassical transport of an exciton wavepacket in the ballistic limit. In reality, one does not excite a single exciton eigenmode but rather a wavepacket of excitons, which may have a Gaussian profile in real-space. In reciprocal space, the time-evolution of such Gaussian wavepacket follows:
\begin{equation}
\label{eq:wavepacketQ}
\Psi(\mathbf{Q},t)=\frac{1}{(2\pi\sigma^{2})^{1/4}}\exp\left[-\frac{\mathbf{Q}^{2}}{4\sigma^{2}}-\frac{i\Omega^{S}_\vec{Q}t}{\hbar}\right],
\end{equation}
where $\Omega^{S}_\vec{Q}$ is the exciton dispersion, and $\sigma$ is the standard deviation of the distribution in reciprocal space. We chart the real-space evolution of the exciton wavepacket by taking the Fourier transform of the reciprocal-space wavepacket.
Fig.~\ref{fig:Fig2} shows the resulting time evolution for the examined excitons in the studied systems. Because we are interested in the initial conditions of the exciton propagation, we look at very short propagation time scales on the order of exciton decoherence times, $\sim$100~fs. Our results show an immediate connection between the dimension-specific exciton bandstructure features around $\textbf{Q}=0$, discussed above, and the wavepacket evolution. 

Naively, we expect a Gaussian wavepacket to remain Gaussian, but this is purely a consequence of the parabolic energy dispersion, which gives the time evolution the general form of a complex Gaussian distribution  (Eq.~\ref{eq:wavepacketQ}). Thus, in all cases, the parabolic exciton bands maintain a Gaussian shape, and for the short time scales considered, they are well approximated as completely non-dispersive. Conversely, the non-parabolic bright exciton bands, which are either discontinuous or non-differentiable at $\textbf{Q}=0$, result in unexpected patterns in the wavepacket time-evolution. In pentacene, the strong anisotropy leads to a cross-like shape evolution of the wavepacket. In MoS$_2$, the linear band forms a ring-like shape, as the wavepacket distribution quickly propagates away from the cusp at $\textbf{Q}=0$, resembling recent experimental observations~\cite{Kulig2018}. A similar effect occurs in black phosphorus, but only along the direction of linear dispersion. Finally, in the (8,0) SWCNT, the non-parabolic band leads to an oscillating pattern as the wavepacket disperses along the nanotube. Thus, even at very short time scales before the onset of diffusive behavior, the initial femtoseconds of the exciton time-evolution can set up an unexpected spatial distribution, which carries over to establish the initial conditions of the exciton diffusion.

In conclusion, we have performed state-of-the-art first principles calculations to realize the relation between exciton bandstructure and materials' symmetry and dimensionality. We show that the long-range exchange interaction gives rise to exciton dispersions that have non-analytic discontinuities in the small-\textbf{Q} limit. In 3D, crystal asymmetry leads to a jump discontinuity in the dipole-allowed exciton bands. In quasi-2D, the exciton has a massless non-analytic dispersion when the exciton center-of-mass momentum has a component parallel to the exciton dipole matrix element. In quasi-1D, there is a removable discontinuity in the dispersion of the bright states at $\textbf{Q}=0$. These exciton dispersions are very different from the commonly used Frenkel and Wannier approximations, where the exciton bandstructure is assumed to be respectively either non-dispersive or parabolic. We show that the non-parabolic exciton dispersion has consequences for initial quasi-ballistic transport of the exciton, leading to unexpected patterns in the time-evolution of an exciton wavepacket, and should thus be considered in a complete picture of exciton dynamics. The detailed understanding of the exciton dispersion developed here can also have consequences for exciton scattering phenomena, including exciton-phonon interaction, Auger processes, and multi-exciton generation.

Acknowledgments: We thank Jeffrey B. Neaton, Felipe H. da Jornada, Alexey Chernikov, and Akshay Rao for valuable discussions. This research was supported by the Center for Computational Study of Excited State Phenomena in Energy Materials (CSEPEM), which is funded by the U.S. Department of Energy, Office of Science, Basic Energy Sciences, Materials Sciences and Engineering Division under Contract No. DE-AC02-05CH11231; by an Israel Science Foundation Grant No.1208/19; and by a Yale/Weizmann Exchange Program Grant. Computational resources of the National Energy Research Scientific Computing Center (NERSC) were used, as well as a PRACE Allocation Grant at the Barcelona Supercomputing Center (BSC-CNS).  
S. R. A. is an incubment of the Leah Omenn Career Development Chair. We acknowledge a research grant from the Peter and Patricia Gruber Awards and an Alon Fellowship.

\noindent
D.Y.Q and S.R.A share equal correspondence of this work.

\bibliography{bibliography}

\end{document}